\documentclass[conference]{IEEEtran}
\IEEEoverridecommandlockouts
\usepackage{cite}
\usepackage{listings}
\usepackage{amsmath,amssymb,amsfonts}
\usepackage{algorithmic}
\usepackage{graphicx}
\usepackage{textcomp}
\usepackage{xcolor}
\usepackage{makecell}
\usepackage{caption}
\usepackage{tikz}
\usepackage{float}
\newfloat{lstfloat}{htbp}{lop}
\floatname{lstfloat}{Listing}
\usepackage[T1]{fontenc}
\def\BibTeX{{\rm B\kern-.05em{\sc i\kern-.025em b}\kern-.08em
    T\kern-.1667em\lower.7ex\hbox{E}\kern-.125emX}}

\newcommand\copyrighttext{%
  \footnotesize For the purpose of open access, the author has applied a Creative Commons Attribution (CC BY) license to any Author Accepted Manuscript version arising}
\newcommand\copyrightnotice{%
\begin{tikzpicture}[remember picture,overlay]
\node[anchor=south,yshift=10pt] at (current page.south) {\fbox{\parbox{\dimexpr\textwidth-\fboxsep-\fboxrule\relax}{\copyrighttext}}};
\end{tikzpicture}%
}
    
\begin{document}

\title{Accelerating stencils on the Tenstorrent Grayskull RISC-V accelerator}

\author{\IEEEauthorblockN{Nick Brown}
\IEEEauthorblockA{\textit{EPCC} \\
\textit{University of Edinburgh}\\
Edinburgh, UK \\
n.brown@epcc.ed.ac.uk}
\and
\IEEEauthorblockN{Ryan Barton}
\IEEEauthorblockA{\textit{Tenstorrent} \\
Toronto, Canada}
}

\maketitle
\copyrightnotice
\begin{abstract}
The RISC-V Instruction Set Architecture (ISA) has enjoyed phenomenal growth in recent years, however it still to gain popularity in HPC. Whilst adopting RISC-V CPU solutions in HPC might be some way off, RISC-V based PCIe accelerators offer a middle ground where vendors benefit from the flexibility of RISC-V yet fit into existing systems.
In this paper we focus on the Tenstorrent Grayskull PCIe RISC-V based accelerator which, built upon Tensix cores, decouples data movement from compute. Using the Jacobi iterative method as a vehicle, we explore the suitability of stencils on the Grayskull e150. We explore best practice in structuring these codes for the accelerator and demonstrate that the e150 provides similar performance to a Xeon Platinum CPU (albeit BF16 vs FP32) but the e150 uses around five times less energy. Over four e150s we obtain around four times the CPU performance, again at around five times less energy.
\end{abstract}


\section{Introduction}

Recent developments, such as large core count commodity available RISC-V CPUs \cite{c920-datasheet} are making RISC-V a more serious proposition for HPC, and indeed benchmarking \cite{brown2023risc} has been encouraging. However, moving wholesale to a RISC-V based CPU system, especially for a supercomputer, is a very significant change, requiring an entirely new hardware and software stack. Whilst there is work going on in all these areas, there is still much to be done to match the support enjoyed by x86 and AArch64.

Instead, an important role that RISC-V might provide for high performance workloads in the short to medium term is as an accelerator. There are numerous RISC-V based PCIe accelerators currently in development, and a major benefit is that these can be fit into existing, non RISC-V, systems. Many of these accelerators have been developed for Artificial Intelligence (AI) and Machine Learning (ML) workloads, and are driven by the current boom in AI. However, fundamentally this hardware provides the ability to accelerate linear algebra operations which is also a fundamental building block of a much wider set of applications, including those in scientific computing. Consequently, there is a role for these accelerator technologies to be leveraged by the HPC community, but a challenge is that often a more flexible programming interface is required compared to providing Tensorflow or Pytorch for ML workloads.

One such RISC-V based accelerator card is the Grayskull developed by Tenstorrent. Available for purchase at a modest price, this commodity card is, as of 2024, one of the few RISC-V based accelerators publicly available. The modest price not only means that these can be leveraged in best-of-class supercomputers, but that they are also suitable for more modest HPC machines and even a realistic proposition as an add on to workstations. However, whilst the Tenstorrent team are making significant progress in supporting general workloads, the Grayskull is most mature for AI inference as this is their primary market. 

In this paper we solve Laplace's equation for diffusion via the Jacobi iterative method, using this as a vehicle to explore accelerating stencil based algorithms on the Tenstorrent Grayskull and their Tensix cores more widely. Stencils are a very common algorithmic pattern in scientific computing \cite{datta2008stencil}, and after describing the Grayskull architecture and Jacobi method in more detail in Section \ref{sec:bg}, we then report the setup used for our experiments throughout this paper in Section \ref{sec:experimental_setup} before describing our initial port to the Grayskull and Tensix cores in Section \ref{sec:initial_jacobi}. Based upon the bottlenecks around data loading and writing identified in Section \ref{sec:initial_jacobi}, we use a streaming benchmark in Section \ref{sec:data_access_strategies} to explore the performance properties of different data access strategies on the Grayskull, drawing general conclusions around how to obtain best performance before leveraging this information to optimise our algorithm in Section \ref{sec:optimised}. We then undertake a performance and energy efficiency comparison between up to four Grayskull cards and a Xeon Platinum CPU in Section \ref{sec:performance}, before drawing conclusions, highlighting recommendations and describing further work in Section \ref{sec:conclusions}. 

The novel contributions of this paper are:

\begin{itemize}
    \item We undertake, to the best of our knowledge, the first study of not only the Tenstorrent Grayskull, but more widely a commodity available RISC-V based PCIe accelerator card, for HPC workloads. 
    \item An exploration of DDR data access strategies on the Grayskull, providing an understanding of best practice around structuring DDR memory access to obtain optimal performance.
    \item Undertaking a performance and energy efficiency analysis of an iterative solver, and the common algorithmic pattern of stencils, on the Grayskull against a server grade CPU.
\end{itemize}

\section{Background}
\label{sec:bg}

RISC-V is an open, community driven, Instruction Set Architecture (ISA) which has been used by a wide range of vendors to build processing technologies. With over 13 billion RISC-V devices manufactured to date, a wide range of hardware companies have begun to leverage this common ISA and benefit from all the work being undertaken in the wider ecosystem. There is a large community involved in progressing RISC-V, ranging from those who are further enhancing the ISA standard to work at the software level, for instance improving compiler support and optimisation of libraries and applications. 

\subsection{Tenstorrent Grayskull}
\label{sec:grayskull}
\begin{figure}[htb]
\centering
 \includegraphics[width=\columnwidth]{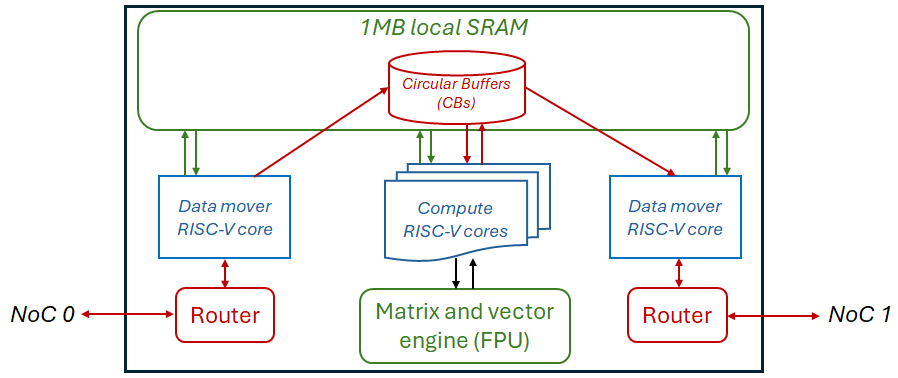}
\caption{A single Tensix core contains five RISC-V \emph{baby cores}, 1MB of SRAM memory, an FPU and two routers.}	
\label{fig:tensix}
\end{figure}

The Grayskull is a PCIe accelerator card developed by Tenstorrent \cite{e150-datasheet}, and whilst newer members of the family, such as the Wormhole, have been announced, the Grayskull by far the most widely available, and the whole family of cards are based upon the same architecture and general design principals. The key components in all these cards are Tenstorrent's Tensix cores, which are illustrated in Figure \ref{fig:tensix}. Tensix cores contain five RISC-V CPUs, known as \emph{baby cores}, a matrix and vector Floating Point Unit (FPU), 1MB of SRAM and two routers each of which are connected to separate Networks on Chip (NoCs). The five RISC-V cores comprise two data mover cores and three compute cores. The data mover cores connect a Tensix core to other Tensix cores, and also to DRAM, moving data into and out of a Tensix core. Programmers typically write two data movement kernels, one for each core, where one core is commonly used for moving data in and the other for moving data out. The data mover cores reside on separate NoCs.

The compute cores drive the matrix and vector engine, known as the FPU, and whilst these operate as three separate cores they are logically viewed by the programmer as one core with a single kernel written and launched upon them which is executed concurrently by each compute core. The compute cores comprise an unpacker core, math core, and packer core however this distinction is only made in the underlying framework which selects specifically which compute core(s) should undertake which operations and this is abstracted from the programmer.

The TT-Metalium framework, \emph{tt-metal}, is Tenstorrent’s low level SDK which exposes direct access to the hardware, providing an API for direct kernel development. A range of ML primitives are built atop this open source framework and then used by Tenstorrent's higher level TT-Buda AI framework. In this work we focus exclusively on tt-metal, using the SDK to develop custom kernels for our Jacobi solver. The SDK provides an API that programmers can use to undertake a range of low level activities such as the movement of data, driving the FPU, and Circular Buffers (CBs). CBs are the way in which RISC-V baby cores in a Tensix core communicate and are First In First Out (FIFO) queues that wrap around. These are split into segments, or pages, and CBs follow a producer-consumer model where one core will add data into the CB and another consume it. The size of each page, along with number of pages is defined by the host code, and the CB producer will call \emph{cb\_reserve\_back} which blocks until a specified number of pages is available in the queue. Once these pages have been filled by the producer, the \emph{cb\_push\_back} API call is issued which will make these available in the CB to the consumer. On the consumer side, \emph{cb\_wait\_front} blocks until a specified number of pages have been made available, or committed to the queue, by the producer and once these have been consumed then \emph{cb\_pop\_front} frees them up so they can be reused by the producer. CBs are a powerful abstraction which provide a pipelined approach between the baby RISC-V cores, enabling these cores to be running concurrently reading in data, computing data and writing out data all on different pages of the CB(s). It is also possible to directly allocate memory in local SRAM. 


The FPU can be viewed as a 16384 bit wide SIMD unit but in addition to supporting basic maths operations such as element wise addition, subtraction, multiplication and division it can also undertake a range of other mathematical and logical operations such a calculating squares, logs, trigonometric functions, conditionals and reductions, as well as higher level operations commonly required for ML workloads such as matrix multiplication, ReLU, sigmoid, and transposition. The FPU in the Grayskull supports a maximum of half precision floating point (both FP16 and BF16), with the Wormhole supporting up to single precision. CBs are provided as arguments to all FPU operations in tt-metal, where the unpacker compute core extracts data, for instance of size 32 by 32 when working with half precision, into tile registers of the FPU, the maths compute core drives the FPU operating upon its tile registers, and then the packer compute core packs the output tile registers into a target CB which can then be consumed by a data mover.

There are two models of the Grayskull, the e75 and e150 with the later providing more resources. In this work we focus on the e150, which contains 120 Tensix cores operating at 1.2 GHz, although only 108 of these are workers (i.e. can be used for compute) and the other 12 are for storage only. The e150 also has 8GiB of DRAM which is split across eight banks, and the card is quoted by Tenstorrent as providing a theoretical peak of 332 FP8 TFLOP/s.

Whilst leveraging the Grayskull for HPC is in its infancy, there have been some early studies of using these cards for ML workloads \cite{thuning2024attention} \cite{doerner2024analysis}, and there is a significant amount of work being undertaken by the vendor to further enhance their SDK. This technology is therefore worth exploring for HPC, not least because it decouples the movement of data from compute where it is possible to be concurrently computing, reading the next tile of data, and writing the previous tile. Furthermore, each Tensix core has 1MB of local SRAM which is a large amount for a cache and, coupled with the ability to write data mover kernels that operate independently and manipulate this memory enables the development of specialised data caching and data reuse approaches that directly suit an application. Indeed, this is one of the major benefits that FPGAs provide to HPC workloads \cite{de2021stencilflow}, where the concurrency provided by an FPGA enables compute to continually operate whilst data is loaded in and out, and it has been found that this is especially beneficial for challenging memory access patterns, such as irregular memory accesses. However, FPGAs are very complicated to program with esoteric tool chains, and by contrast the Grayskull is far simpler because it is built around CPU cores. Consequently, it is interesting to understand whether the Grayskull, and Tensix cores more widely, can provide similar memory specialisation benefits as FPGAs, but in a more programmable manner.

\subsection{Jacobi iterative method}
\label{sec:jacobi}
Jacobi's algorithm is the simplest iterative solution method. However, whilst the convergence rate of this algorithm is inferior to other, more complex methods, the memory access patterns represent these more complex methods, and also a much wider set of stencil based codes, with the simplicity enabling us to focus on the underlying optimisations for the Grayskull.

When using Jacobi, for a linear system, $Ax=b$, one starts with a trial solution $x_{0}$ and generates new solutions iteratively, according to $x_{i}^{(k)} = \frac{1}{a_{ii}}(b_{i}-\sum_{i != j}a_{ij}x_{j}^{(k-1)})$ where $k$ is the iteration number. The algorithm terminates once a fixed number of iterations have been completed. In this paper we solve Laplace's equation for diffusion, $\bigtriangledown^{2}u=0$ in two dimensions using a five point stencil. Listing \ref{lst:jacobi-pseudo} provides a pseudo code sketch of this algorithm where there are two arrays, \emph{unew} and \emph{u}. At each iteration, the value calculated for every grid point is the average of its neighbouring values and this is stored in \emph{unew}. The \emph{unew} and \emph{u} arrays are separate so there are no data races, because data being read was calculated in the previous iteration. At the end of an iteration the \emph{unew} and \emph{u} arrays are swapped.

\begin{lstlisting}[frame=lines,  label=lst:jacobi-pseudo, language=matlab, caption=Pseudo code of Jacobi iterative method solving Laplace's equation for diffusion in two dimensions]
for all iterations:
    for all grid points i and j:
	   unew(i,j) = 0.25 * ( u(i+1,j)+u(i-1,j)+ u(i,j+1)+u(i,j-1) )
    swap unew and u
\end{lstlisting}

\begin{figure}[htb]
\centering
 \includegraphics[width=0.75\columnwidth]{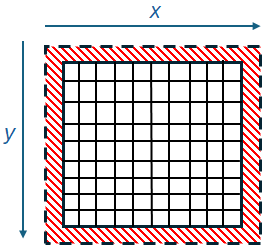}
\caption{Illustration of a domain surrounded by boundary conditions for stencil based computation.}	
\label{fig:laplace}
\end{figure}

The Jacobi iterative method is an example of a stencil based algorithm, where values from neighbouring grid cells are required during the calculation of the current cell's value. For Laplace's equation for diffusion we have a stencil depth of one, which means that only one neighbouring value in each dimension is required. On the edges additional grid cells are required which are called the halos, and these serve as neighbours to the grid cells which are left, right, top or bottom most in the domain. At the global domain level these are fixed boundary conditions, whereas if a chunk is taken to be processed locally, for instance when decomposing the domain across processing elements to run in parallel, these halos must also be included and represent the stencil depth number of values from neighbouring chunks. An illustration of this is provided in Figure \ref{fig:laplace}, where the domain of grid cells is surrounded by the boundary conditions in red. In Laplace's equation for diffusion these boundary conditions vary from one side to the other, for instance on the left might be high values and the right low values. At the start of the algorithm values in each grid cell are set to an initial guess, often zero or one, and then from one Jacobi iteration to the next the boundary condition values propagate, or diffuse, through the system until a stable state is reached after many iterations which represents the final solution.

Stencils are a common algorithmic pattern in scientific computing and underlie many HPC applications including atmospheric modelling \cite{brown2020highly}, Computational Fluid Dynamics (CFD) \cite{jamshed2015using}, and seismology \cite{luporini2020architecture}. Whilst the plus one and minus one in the contiguous memory dimension is straightforward for CPUs to cache and prefetch, the offsets in the non-contiguous dimension are more difficult. Indeed, FPGAs have proven effective for stencil based algorithms by leveraging a shift buffer \cite{brown2021accelerating}. This is a bespoke caching mechanism which stores and serves previously read data until it is no longer required, avoiding duplicate reads. Consequently, understanding how to best represent this common algorithmic pattern on the Grayskull is not only very topical to HPC, but furthermore acts as an interesting case study around whether the data mover cores can provide effective bespoke memory management.

\section{Experimental setup}
\label{sec:experimental_setup}
Results reported from the experiments run throughout this paper are averaged over five runs. All Grayskull codes are run on an e150, hosted in a machine by Tenstorrent, connected to the main board by PCIe Gen 4. This machine contains two 24-core AMD EPYC 7352 CPUs and 256GB of DRAM. All experiments are built with version 0.50 of the tt-metal framework, and Clang 17 is used to compile host codes. We build and execute CPU codes on a 24-core 8260M Cascade Lake Xeon Platinum CPU, which is equipped with 512GB of DRAM and codes are compiled using GCC version 11.2. Multi-core codes on the CPU are multi threaded using OpenMP. Energy usage on the CPU is based upon values reported by RAPL, and on the e150 from the Tenstorrent System Management Interface (TT-SMI).

All results reported on the e150 are running in BF16, whereas on the CPU they are single precision floating point. Whilst this is not a perfect comparison, it is the highest precision supported by the e150 and lowest supported by the CPU. Unless otherwise stated, the results reported for the Grayskull include the overhead of transferring data to and from the card over PCIe.

\section{Initial Jacobi implementation}
\label{sec:initial_jacobi}

\begin{figure}[htb]
\centering
 \includegraphics[width=\columnwidth]{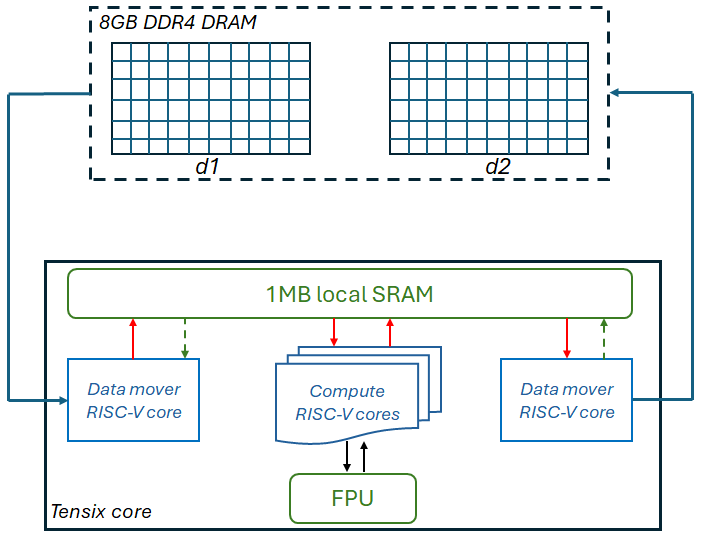}
\caption{Initial design, where a Tensix core retrieves data from DRAM, serves it to the compute cores which drive the FPU, and results are then written back to DRAM.}	
\label{fig:overall_approach}
\end{figure}

Figure \ref{fig:overall_approach} illustrates our initial overarching design for leveraging a Tensix core for this application, where one of the data mover cores reads input data from DRAM and stores this in the Circular Buffers (CBs) that are held in local SRAM. These CBs are then made available to the compute cores, which unpack the data to their tile registers, undertake mathematical operations on the FPU, and then pack the results in the tile registers to a CB. This output CB is then consumed by the other data mover core which writes the results back to DRAM. As described in Section \ref{sec:jacobi}, the algorithm reads from array \emph{u} and writes to \emph{unew} and these are swapped between iterations. In our approach, the data mover cores track the iteration number, and depending upon whether the iteration number is even or odd selects the mapping between the \emph{d1} and \emph{d2} data areas in Figure \ref{fig:overall_approach} to the \emph{u} and \emph{unew} arrays, effectively cycling between these from one iteration to the next.

The green dashed lines between the data mover cores and local SRAM represent a semaphore where, for consistency, the data mover core that is reading blocks on a semaphore which the data mover core that is writing updates to ensure that it can move to the next iteration. 

\subsection{Compute kernel design}

\begin{figure}[htb]
\centering
 \includegraphics[width=0.75\columnwidth]{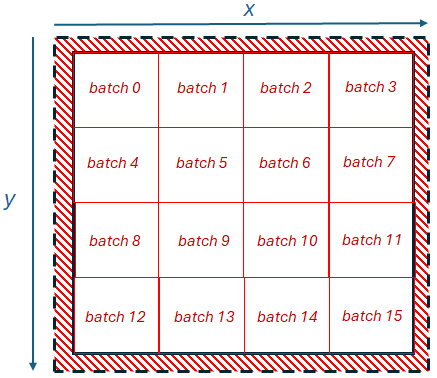}
\caption{Illustration of decomposing the domain into distinct batches of size 32 by 32 BF16 elements.}
\label{fig:decomp_domain}
\end{figure}

The matrix and vector FPU engine, fed by the compute cores in Figure \ref{fig:overall_approach} works on chunks of data that are 16384 bits wide. The FPU in the Grayskull supports at most half precision, with all numbers in our code bfloat16 (BF16), and-so the FPU computes on 1024 BF16 elements at a time. This results in a tile of size 32 by 32 BF16 numbers, and Figure \ref{fig:decomp_domain} illustrates splitting the 2D domain up into batches of this size, representing the same domain illustrated in Figure \ref{fig:laplace}, but each batch containing 32 by 32 grid cells.

As sketched in Listing \ref{lst:jacobi-pseudo}, the value for each grid cell is calculated as the average of its neighbouring values. Consequently, there are four tiles of size 32 by 32 elements generated from each batch. The first and second tiles represent values which are offset by minus one and plus one in the X dimension respectively, and the third and fourth tiles are offset by minus one and plus one in the Y dimension respectively. These tiles are packed into four separate CBs by the data mover and provided to the compute cores. Listing \ref{lst:compute} illustrates the code running on the compute cores for a single iteration and single batch, which operates on these tiles, undertaking the addition of four neighbouring values and multiplication by 0.25 to obtain the average. For brevity we have omitted calling the initialisation functions and tile register acquire routines.

\begin{lstlisting}[frame=lines, label=lst:compute, caption=Compute kernel code driving the FPU based on four tiles per batch.]
constexpr uint32_t dst0 = 0;

cb_wait_front(cb_in0, 1);
cb_wait_front(cb_in1, 1);
add_tiles(cb_in0, cb_in1, 0, 0, dst0);
cb_pop_front(cb_in1, 1);
cb_pop_front(cb_in0, 1);
    
cb_reserve_back(cb_intermediate, 1);
pack_tile(dst0, cb_intermediate);
cb_push_back(cb_intermediate, 1);
    
cb_wait_front(cb_in2, 1);
cb_wait_front(cb_intermediate, 1);
add_tiles(cb_in2, cb_intermediate, 0, 0, dst0);
cb_pop_front(cb_intermediate, 1);
cb_pop_front(cb_in2, 1);   

cb_reserve_back(cb_intermediate, 1);
pack_tile(dst0, cb_intermediate);
cb_push_back(cb_intermediate, 1);

// Undertaking the same addition for the third CB

cb_wait_front(cb_intermediate, 1);
mul_tiles(cb_scalar, cb_intermediate, 0, 0, dst0);
cb_pop_front(cb_intermediate, 1);

cb_reserve_back(cb_out0, 1);
pack_tile(dst0, cb_out0);
cb_push_back(cb_out0, 1);
\end{lstlisting}

Lines 3 and 4 in Listing \ref{lst:compute} block for a tile to be available in the \emph{cb\_in0} and \emph{cb\_in1} circular buffers, which corresponds to the \emph{i-1} and \emph{i+1} tiles. An element wise addition of these tile values is undertaken by the FPU at line 5, with the result stored in the zero set of tile registers (\emph{dst0}). Lines 6 and 7 then free up these pages in the CBs, so the data loading core can reuse this area. In total, we allocate four pages for each CB meaning that data loading and compute can overlap.

Line 9 of Listing \ref{lst:compute} reserves a page in the \emph{cb\_intermediate} CB, before the data in the \emph{dst0} register is packed into this CB at line 10 and this is then made available to the consumer at line 11. The consumer of this intermediate CB is in fact the same compute kernel because this CB is used in the subsequent maths operations as an input. As described in Section \ref{sec:grayskull}, all maths operations take CBs as inputs, and-so in order to multiply tile values by the 0.25 constant scalar value a CB must be provided where all 1024 values are 0.25. This is \emph{cb\_scalar} in Listing \ref{lst:compute}, which is a CB filled by a data mover core on program initialisation, with the compute core issuing \emph{cb\_reserve\_back} also on initialisation. Lastly, at lines 29 to 31, a tile is reserved in the \emph{cb\_out0} CB, the tile registers are packed into this and at line 31 this is then made available to the consumer which is the data mover that is writing result data back to DRAM.

As an aside, it was our hypothesis that operating upon 2D tiles might potentially offer a performance advantage because it promotes increased data reuse. When considering this algorithm executing in a scalar fashion on the CPU, as per Listing \ref{lst:jacobi-pseudo}, the majority of grid points must be read four times per iteration because, as the algorithm works through the grid, most data elements are used in all four locations of \emph{u}. However, because the Tensix core is working in tiles of 32 by 32 BF16 elements, many of these replicated accesses reside within the same tile and the data is already present. Consequently, only the outer 126 elements are required as halos by a different tile, and in that case are only required once. Therefore, working in these blocks has the potential to more naturally provide date reuse.

\subsection{Data movement approach}

Each batch of data requires not only the 32 by 32 grid values, but also the halos. Therefore, each batch requires 34 non-contiguous reads from memory, each of which are of size 34 BF16 elements, or 68 bytes. Our initial approach for data reading is illustrated in Listing \ref{lst:initial-dataread}, which iterates through the rows in the Y dimension for a specific batch. Line 2 calculates the address offset, where \emph{batch\_offset} has already been calculated for each batch to offset it in the X and Y dimension and is omitted for brevity. Line 3 obtains the address to access on the NoC that will resolve to the correct location in DRAM, with \emph{noc\_x} and \emph{noc\_y} providing the location of the DRAM bank on the NoC. Line 4 then issues the reading of data from this address, and numbers of elements in Listing \ref{lst:initial-dataread} are multiplied by two to convert them into bytes. Reads are non blocking, and once these are issued across the entire batch a \emph{noc\_async\_read\_barrier} call is made which blocks until all reads have been completed. At this point, the 34 by 34 tile of elements is held in the local SRAM buffer and this is then copied to the four CB, each of size 32 by 32, extracting the appropriate data for each tile with the corresponding halos and offsets into local memory applied. The writing of results is simpler, as there is only one output CB from the compute cores per batch and this is already of size 32 by 32 elements, which can be written directly to DRAM.

\begin{lstlisting}[frame=lines,  label=lst:initial-dataread, caption=Initial approach for reading data from DRAM.]
for (uint32_t j=0;j<BATCH_SIZE_IN_Y;j++) {
  std::uint32_t addr_offset=(j*total_size_in_x)+batch_offset;
  uint64_t noc_addr = get_noc_addr(noc_x, noc_y, ddr_addr+(addr_offset*2));
  noc_async_read(noc_addr, local_buffer+(j*(BATCH_SIZE_IN_X)*2), 34*2);
}
noc_async_read_barrier();
// Issue memory copies to four CBs based on the local buffer 
\end{lstlisting}

However, upon developing this approach we found that it resulted in incorrect values starting from the second row of Y downwards. Whilst there were no compile or runtime errors reported, from experimentation we found that all DRAM accesses must be aligned on 256 bit boundaries, and any which are unaligned provide incorrect values when reading data and corrupt values being stored when writing. Even though our domain sizes tend to be to the power of two, because we have boundary conditions on the left and right, after the first read subsequent reads are unaligned.

For reading data we adopted the approach shown in Listing \ref{lst:align-dataread}, where the \emph{address} argument is the address to read from and \emph{start\_address} is the starting address of the data in DRAM (which is always aligned). Line 2 calculates the number of bytes that the read is unaligned by and stores this in \emph{offset}, with this then being used at line 3 to adjust the starting read location, working backwards to align the read and storing this in \emph{offset\_start}. The \emph{read\_size} variable at line 4 is the number of bytes to read including the offset and the NoC address is retrieved at line 6, with the read itself undertaken at line 7, storing into the local buffer. The additional offset that was read is returned back to the caller and the caller can then use this to unpack the local buffer starting from this offset and ignore the additional preliminary data that was also read to ensure alignment.

\begin{lstlisting}[frame=lines,  label=lst:align-dataread, caption=Approach for reading data from DRAM to ensure alignment.]
std::uint32_t read_data(std::uint32_t address, std::uint32_t starting_address, std::uint32_t noc_x, std::uint32_t noc_y, std::uint32_t size, std::uint32_t buffer_addr) {
    std::uint32_t offset=(address - starting_address) % ALIGNMENT;
    std::uint32_t offset_start=address-offset;
    std::uint32_t read_size=size+offset;

    uint64_t noc_addr = get_noc_addr(noc_x, noc_y, offset_start);
    noc_async_read(noc_addr, buffer_addr, read_size);
    noc_async_read_barrier();
    return offset;
}
\end{lstlisting}

We found that this approach worked well for reading data, providing consistent results irrespective of the starting location in memory that was being read from. However, a similar approach for writing data did not work. We calculated the additional number of elements that had to be written, read these from DDR and packed them into a temporary buffer with the rest of the data which was then written to DRAM. However, this resulted in corrupted values in DRAM, which was likely because there is no guarantee between the ordering of data reads and writes. Consequently, if values have been recently updated then reading these same values as the additional values from DRAM could result in stale values. We found that contiguous data writes of unaligned data does work as long as these come from separate locations in a buffer which are not overwritten, and indeed in most cases it was possible to use a small number of buffers across writes and cycle between them. From this we suspect that the DRAM controllers are undertaking some merging of data writes to handle the unaligned case. However, this approach did not work for non-contiguous data writes, and that is the memory access pattern required by our code due to batches of 32 by 32 elements as illustrated in Figure \ref{fig:decomp_domain}.

\begin{figure}[htb]
\centering
 \includegraphics[width=0.75\columnwidth]{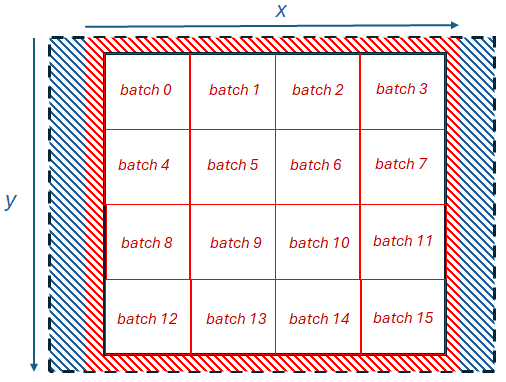}
\caption{Illustration of additional 256 bit wide allocation on the left and right of the domain, containing empty values apart from the boundary conditions so that writing of 32 by 32 result tiles is always aligned.}	
\label{fig:align_domain}
\end{figure}

Consequently, to ensure that our data writes were always aligned we limited the domain size to a power of two, and allocate  an initial 256 bit wide area of memory on the left of the domain, and another allocated on the right. This is illustrated in Figure \ref{fig:align_domain}, where these new values are mostly empty (in blue) apart from the boundary conditions that occupy 2 bytes each. 

\subsection{Initial performance}
Based upon the design detailed in this section we then undertook performance experimentation and tuning on one Tensix core using a problem size of 512 by 512 BF16 elements and 10000 iterations. The results of this experiment are reported in Table \ref{tab:initial-perf} and measured in billion points processed per second (GPt/s), where \emph{CPU single core} is the reference algorithm running over a single core of the Xeon Platinum Cascade Lake CPU. \emph{Initial} is the initial version of our code on the e150 we have described in this section, then optimising the writing of data to issue the \emph{noc\_async\_write\_barrier} write synchronisation at the batch level rather than for each individual write request, which resulted in a modest performance improvement.

\begin{table}[htb]
    \centering
    \caption{Performance of Tensix core executing Jacobi solver using problem size 512 by 512 (262144 BF16 elements) over 10000 iterations.}
    \label{tab:initial-perf}
    \begin{tabular}{|c|c|}
    \hline           
     \textbf{Version} & \textbf{Performance (GPt/s)} \\
      \hline
    CPU single core & 1.41 \\
    \hline
    Initial & 0.0065\\
    Data write optimised & 0.0072\\
    Double buffering & 0.0140\\
    \hline
    \end{tabular}
\end{table}

A more substantial performance improvement was obtained on the e150 by double buffering the reading of data in the data mover core. In this approach we block for outstanding reads only at the start of a batch, and then issue calls to retrieve data for the next batch into the next buffer in local SRAM. Whilst these reads are on-going, memory copies are undertaken to copy data into the four CBs from the current batch held in the current buffer, before iterating onto the next batch. The expectation here was that this would provide overlapping of data reading and memory copying from local buffers to CBs. This approach was worthwhile, and delivered around twice the performance for our code, however the Tensix core was still around 100 times slower than the CPU core. Incidentally, we found that enabling the print server, which enables messages to be printed by the Tensix cores, incurred significant overhead and-so whilst this was useful during development it was disabled for all production runs.

To understand where the bottlenecks in our design lay we deactivated selected parts of our design and retimed. This involved being able to selectively switch on and off the data loading, compute and data writing, whilst keeping the CB structure and synchronisation between the data mover and compute cores. The objective was to see what parts of the code made the biggest impact on overall runtime, and the results of this experiment are reported in Table \ref{tab:strip-perf}. It can be seen that without any data reading, writing or compute the performance is 7.574 GPt/s which is well in excess of that provided by the CPU core. Enabling only the compute component resulted in performance of 1.387 GPt/s, and at that point we experimented with different compute kernel designs such as initialising the maths addition operators to accumulate using values held in the destination registers to avoid some packing and unpacking of CBs, but this actually resulted in lower performance. We therefore concluded that, given the structure of compute required by this algorithm, 1.387 GPt/s which is comparable to the performance delivered by a CPU core, is the realistic maximum to aim for.

\begin{table}[htb]
    \centering
    \caption{Performance of a single Tensix core executing Jacobi solver using problem size 512 by 512 (262144 BF16 elements) over 10000 iterations when disabling specific components.}
    \label{tab:strip-perf}
    \begin{tabular}{|cccc|c|}
    \hline           
     \textbf{Read} & \textbf{Memcpy} & \textbf{Compute} & \textbf{Write} & \textbf{Performance (GPt/s)} \\
      \hline   
    N & N & N & N & 7.574\\    
    N & N & Y & N & 1.387\\
    N & N & N & Y & 0.278\\
    Y & N & N & N & 0.205\\
    N & Y & N & N & 0.014\\
    Y & Y & N & N & 0.013\\        
    \hline
    \end{tabular}
\end{table}

However, our code was obtaining no where near this 1.387 GPt/s level of performance and it can be seen from Table \ref{tab:strip-perf} that the major bottleneck was in the movement of data. When enabling only data writing, performance dropped considerably to 0.278 GPt/s, which is slightly faster than only enabling data reading. It was found that the greatest bottleneck was in the copying of memory by the data mover core, copying data that has been read into a local buffer into the CBs. Whilst we had assumed that double buffering would help hide the overhead of this, and it did provide a modest performance improvement, clearly this was not sufficient and there was still significant stalling. Furthermore, even if double buffering did entirely hide the overhead of memory coping, we would still have the overhead of reading data and based on the results in Table \ref{tab:strip-perf} the best that we could hope for would be 0.205 GPt/s which is around 7 times slower than a core of the CPU.

\section{Exploring data access strategies}
\label{sec:data_access_strategies}

Based upon the results reported in Table \ref{tab:strip-perf} it was clear that data movement in our code required significant redesign. However, it was not clear which was the best strategy to adopt, and-so in order to inform this choice we undertook a series of performance experiments using a streaming benchmark. All results reported in this section are kernel execution time only and do not include data transfers to or from the card. 

This streaming benchmark loads integers from DRAM as quickly as possible by one data mover core, passes these onto the other data mover core which writes them back to DRAM as quickly as possible. Throughout this subsection we use a problem size of 4096 by 4096 32-bit integers, and this enabled us to first experiment with different read chunk sizes to understand the performance implications of issuing fewer, larger reads and writes compared to more frequent, smaller DRAM memory accesses. 

\begin{table}[htb]
    \centering
    \caption{Runtime comparison for streaming benchmark with a problem size of 4096 by 4096 32-bit integers, memory accesses are contiguous and the batch size of memory accesses is varied, with and without synchronisation after each access.}
    \label{tab:contig_chunk}
    \begin{tabular}{|c|c|cc|cc|}
    \hline  
    \textbf{Batch size} & \textbf{DRAM} & \multicolumn{2}{c|}{\textbf{Read Runtime (s)}} & \multicolumn{2}{c|}{\textbf{Write Runtime (s)}}\\
      \textbf{(bytes)} & \textbf{requests / row} & \textbf{no sync} &  \textbf{sync} & \textbf{no sync} &\textbf{sync}\\
      \hline
    16384 & 1 & 0.011 & 0.011 & 0.011 & 0.011 \\
    8192 & 2 & 0.011 & 0.011 & 0.011 & 0.016 \\
    4096 & 4 & 0.012 & 0.013 & 0.011 & 0.020 \\
    2048 & 8 & 0.012 & 0.020 & 0.011 & 0.023 \\
    1024 & 16 & 0.016 & 0.034 & 0.011 & 0.031 \\
    512 & 32 & 0.031 & 0.074 & 0.011 & 0.038 \\
    256 & 64 & 0.039 & 0.201 & 0.011 & 0.053 \\
    128 & 128 & 0.067 & 0.327 & 0.014 & 0.093 \\
    64 & 256 & 0.122 & 0.802 & 0.027 & 0.182 \\
    32 & 512 & 0.238 & 1.571 & 0.052 & 0.360 \\
    16 & 1024 & 0.470 & 3.150 & 0.104 & 0.718 \\
    8 & 2048 & 0.916 & 6.331 & 0.206 & 1.436 \\
    4 & 4096 & 1.761 & 12.659 & 0.411 & 2.873 \\
    \hline
    \end{tabular}
\end{table}

Table \ref{tab:contig_chunk} reports results from this experiment, where we accessed each 4096 row one after the other and varied the read chunk size for data within each row. The maximum batch size is 16384, meaning that all 4096 integers in a row are read, or written, in one memory access, and for example 8192 means that a row will be accessed using two requests. We experimented with the impact of the batch size for reading and writing, and whilst the reading experiments were being undertaken then the batch size was fixed as 16384 for writing and vice versa. We also experimented with synchronising on a batch by batch basis, \emph{sync} in Table \ref{tab:contig_chunk}, where each memory access is followed immediately by the blocking call to wait for its completion, and also only synchronising at the row level, \emph{no sync} in Table \ref{tab:contig_chunk}, when all memory accesses for the row are issued before blocking on their completion. 

It can be observed in Table \ref{tab:contig_chunk} that when synchronising only at the row level, down to a chunk size of 1024 bytes there is little difference in reducing the batch size, however beyond this performance starts to degrade significantly. For the synchronous approach, blocking after each memory access, performance starts to degrade from a batch size of 4096 bytes when reading, illustrating the additional cost of excessive synchronisation. Interestingly, the impact of the batch size, with or without synchronisation, is far greater for reading than it is for writing.

\begin{table}[htb]
    \centering
    \caption{Runtime comparison non-contiguous streaming benchmark with a problem size of 4096 by 4096 32-bit integers, memory accesses are non-contiguous and the batch size of memory accesses is varied, with and without synchronisation after each access.}
    \label{tab:noncontig_chunk}
    \begin{tabular}{|c|c|cc|cc|}
    \hline  
    \textbf{Batch size} & \textbf{DRAM} & \multicolumn{2}{c|}{\textbf{Read Runtime (s)}} & \multicolumn{2}{c|}{\textbf{Write Runtime (s)}}\\
      \textbf{(bytes)} & \textbf{requests / row} & \textbf{no sync} &  \textbf{sync} & \textbf{no sync} &\textbf{sync}\\
      \hline
    16384 & 1 & 0.011 & 0.011 & 0.011 & 0.011 \\
    8192 & 2 & 0.011 & 0.011 &  0.011 &  0.014 \\
    4096 & 4 & 0.012 & 0.012 & 0.011 & 0.020 \\
    2048 & 8 & 0.013 & 0.021 & 0.011 & 0.021 \\
    1024 & 16 & 0.016 & 0.042 & 0.012 & 0.029 \\
    512 & 32 & 0.031 & 0.077 & 0.017 & 0.032 \\
    256 & 64 & 0.042 & 0.201 & 0.022 & 0.052 \\
    128 & 128 & 0.082 & 0.340 & 0.040 & 0.095 \\
    64 & 256 & 0.148 & 0.809 & 0.074 & 0.182 \\
    32 & 512 & 0.275 & 1.597 & 0.143 & 0.361 \\
    16 & 1024 & 0.544 & 3.219 & 0.280 & 0.721 \\
    8 & 2048 & 1.081 & 6.491 & 0.556 & 1.441 \\
    4 & 4096 & 1.969 & 13.013 & 0.715 & 2.882 \\
    \hline
    \end{tabular}
\end{table}

We then repeated this experiment, but accessed data in a non-contiguous fashion where each batch proceeds downwards through the Y dimension, and-so it is guaranteed that subsequent memory accesses are non-contiguous. The results of this experiment are reported in Table \ref{tab:noncontig_chunk} where it can be observed that there is a small to medium performance impact when accessing data in a non-contiguous fashion compared to when the data is contiguous, and this is especially the case as the batch size is reduced. This second experiment more closely represents the memory pattern of our Jacobi code, which reads 34 non-contiguous chunks of 68 bytes for each batch. We repeated the experiment for a variety of different sizes in Y, and found that for all of these experiments performance started to degrade at around a batch size between 1024 to 512 bytes. This therefore suggests that the bottleneck is not necessarily the number of requests on the NoC, but instead the width of DRAM access, with the Grayskull DMA engines and DRAM controllers seeming to favour accesses with larger widths.

We then repeated the contiguous experiment, reading and writing contiguously with a batch size of 16384 bytes. However instead of receiving into the CB directly, instead we read data into a local buffer and then after blocking for all memory accesses in the row to complete issued a memory copy to copy this data into the CB. This resulted in a runtime of 0.106 seconds, which is around ten times slower than when reading directly into the CB and illustrates the overhead involved in copying data as part of the data access strategy. This confirms the observation in Table \ref{tab:strip-perf} that the greatest overhead in our Jacobi code was to be found in the memory copying from local buffers into the four CBs.

\begin{table}[htb]
    \centering
    \caption{Runtime comparison for streaming benchmark with a problem size of 4096 by 4096 32-bit integers. Each memory read is replicated by a specific factor to understand the impact of replicating data reads.}
    \label{tab:replication}
    \begin{tabular}{|c|c|}
    \hline  
    \textbf{Replication factor} & \textbf{Runtime (s)}\\
      \hline
    1 & 0.011\\
    2 & 0.017\\
    4 & 0.033\\
    8 & 0.055\\
    16 & 0.098\\
    32 & 0.185\\
    \hline
    \end{tabular}
\end{table}

The overhead to be found in undertaking memory copying on the data mover core illustrates that, for performance, one should avoid reading data into a local buffer and then copying this into CB(s). However, we adopted this approach in our Jacobi code because each CB required similar data but at slightly different locations in the grid due to the offsets. The alternative would be to undertake four separate reads from DRAM, where each reads directly into the corresponding CB based upon the offset applied to the DRAM memory location. There would be an additional complexity here, were our algorithm for unaligned memory accesses in Listing \ref{lst:align-dataread} reads additional data at the start, and it would be difficult to handle this using the CBs as that additional data would need to be removed. Irrespective, it was instructive to explore the performance properties of issuing replicated reads in this fashion and-so, we undertook an experiment using the same benchmark and domain size, with a batch size of 16384 bytes, where each data access is replicated to also read in the \emph{n} previous rows held in DRAM. The results of this experiment are reported in Table \ref{tab:replication} and it can be seen that even adding an additional single read results in overhead, which increases quickly with the number of additional replicated reads. This demonstrates that we need to both avoid memory copies and additional DRAM reads to deliver optimal performance.

Until this point we have allocated DRAM all in a single bank, however the tt-metal SDK provides the ability to interleave memory across the banks. The e150 contains eight DDR banks, and-so splitting up memory across these could potentially help alleviate pressure on the memory subsystem. Tt-metal cycles pages across banks, with a page size of up to 64KB supported. Table \ref{tab:page_size} reports performance of our same streaming benchmark and problem size, where we varied the page size and replicated memory accesses. The first row, \emph{none}, represents the existing approach without interleaving. It can be seen that, with no replication of memory accesses, there is no performance benefit in interleaving memory. However, when we replicate memory accesses there is a significant benefit, for instance with a page size of of 32KB or 16KB performance with a replication factor of 32 is double that when allocating the entire domain into a single bank. This demonstrates that there is no real downside to using memory interleaving as long as the page size is set appropriately, and when the DDR is under high load it can improve performance considerably.

\begin{table}[htb]
    \centering
    \caption{Runtime comparison for streaming benchmark with a problem size of 4096 by 4096 32-bit integers, running with different page sizes across replication factors.}
    \label{tab:page_size}
    \begin{tabular}{|c|cccc|}
    \hline  
    \textbf{Page size} & \multicolumn{4}{c|}{\textbf{Runtime (s) with replication factor}}\\
      \textbf{(bytes)} & \textbf{0} & \textbf{8} & \textbf{16} & \textbf{32} \\
      \hline
    none & 0.010 & 0.047 & 0.086 & 0.162 \\
    64K & 0.013 & 0.034 & 0.050 & 0.084 \\
    32K & 0.012 & 0.030 & 0.046 & 0.079 \\
    16K & 0.013 & 0.030 & 0.046 & 0.079 \\
    8K & 0.015 & 0.042 & 0.072 & 0.131 \\
    4K & 0.015 & 0.075 & 0.136 & 0.258 \\
    2K & 0.021 & 0.148 & 0.274 & 0.527 \\
    1K & 0.038 & 0.302 & 0.565 & 1.094 \\
    \hline
    \end{tabular}
\end{table}

We then undertook this same experiment, but with no data access replication, and scaled the number of Tensix cores which were decomposed vertically in the Y dimension. Table \ref{tab:page_size_scaling} reports the results of this experiment and surprisingly this does not scale beyond two Tensix cores, irrespective of the page size, which suggests that we are running out of NoC and/or DDR bandwidth. Considering that this is a streaming style benchmark, with no compute, this places considerable bandwidth pressure on the NoC and DDR, so it is not overly surprising but does potentially illustrate a limitation when we come to scaling our Jacobi solver code. Overall, from Table \ref{tab:page_size_scaling} it can be seen that there is no benefit in using interleaving, although smaller page sizes do provide improved scaling but the initial overhead is also greater.

\begin{table}[htb]
    \centering
    \caption{Runtime comparison for streaming benchmark with a problem size of 4096 by 4096 32-bit integers, running with different page sizes across different numbers of Tensix cores}
    \label{tab:page_size_scaling}
    \begin{tabular}{|c|cccc|}
    \hline  
    \textbf{Page size} & \multicolumn{4}{c|}{\textbf{Runtime (s) with number of Tensix cores}}\\
      \textbf{(bytes)} & \textbf{1} & \textbf{2} & \textbf{4} & \textbf{8} \\
      \hline
    none & 0.010 & 0.005 & 0.005 & 0.005 \\
    64K & 0.011 & 0.006 & 0.007 & 0.007 \\
    32K & 0.012 & 0.005 & 0.007 & 0.007 \\
    16K & 0.013 & 0.006 & 0.007 & 0.007 \\
    8K & 0.015 & 0.010 & 0.007 & 0.007 \\
    4K & 0.015 & 0.008 & 0.005 & 0.005 \\
    2K & 0.021 & 0.010 & 0.006 & 0.007 \\    
    \hline
    \end{tabular}
\end{table}

We therefore conclude several lessons learnt from the experiments in this section:
\begin{itemize}
    \item Fewer, larger, DRAM accesses are generally favoured and there is overhead imposed by many small accesses.
    \item Contiguous DRAM accesses generally provide better performance than non-contiguous DRAM accesses.
    \item There is a considerable overhead involved in memory copying between CBs and local buffers.
    \item There are overheads associated with additional DRAM memory accesses, for instance replicating previous reads, but this is somewhat ameliorated by interleaving.
\end{itemize}

\section{Optimised Jacobi kernel}
\label{sec:optimised}
\begin{figure}[htb]
\centering
 \includegraphics[width=0.75\columnwidth]{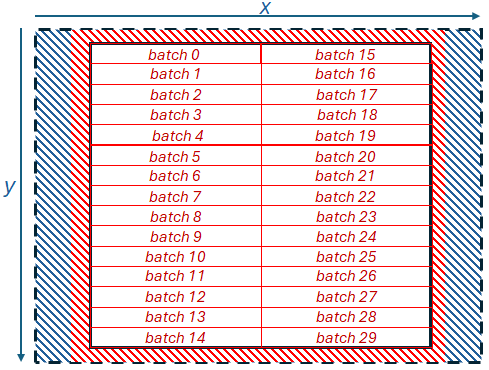}
\caption{Illustration of decomposing the domain into distinct batches of size 1024 elements along the X dimension only}	
\label{fig:revised_decomp_domain}
\end{figure}

Based upon the lessons learnt in Section \ref{sec:data_access_strategies}, we redesigned our Jacobi kernel to remove memory copies and avoid replicated memory accesses. Instead of working in square tiles of 32 by 32 elements as per Section \ref{sec:initial_jacobi}, we modified our code to operate in one dimension chunks of 1024 BF16 elements, 2048 bytes, in order to read data contiguously for each tile in one large read. This is illustrated in Figure \ref{fig:revised_decomp_domain}, which follows the same general approach for alignment but now our algorithm works downwards in the Y dimension reading 1026 elements for each batch, which is the batch's 1024 elements plus two halos on either side. The compute kernel requires the current batch, along with the previous (upper) and next batch (lower). To avoid duplicate reading of data, we allocate enough memory in the core's local memory buffer for four batches and when working in a column of batches in the Y dimension we read batches 0 and 1 and 2 immediately. Then, starting at the first batch (batch zero in Figure \ref{fig:revised_decomp_domain}) we synchronise memory reads immediately, issue a non-blocking read for two batches ahead (batch 2 in Figure \ref{fig:revised_decomp_domain}) and make available to the compute cores data that has been read for the current batch, the previous batch and the next batch.

Whilst this approach results in fewer, larger, memory reads that are contiguous in nature, because we are reading into local SRAM buffers, copying into the CBs is still required and as highlighted in Section \ref{sec:data_access_strategies} this is very expensive. Based upon the current tt-metal API this is inevitable, however because tt-metal is open source we were able to explore the implementation of CBs and discovered that each CB is represented by a structure which has \emph{fifo\_rd\_ptr} and \emph{fifo\_wr\_ptr} fields that point to the memory that the CB will next read from and write to respectively. We can modify the \emph{fifo\_rd\_ptr} field to instead point to a different location in local memory and when this CB is provided to the maths operation as an argument, it is this data which is read. The fact that we are reading in rows of 1024 elements is crucial here for the plus or minus one in the X dimension, because as we only have two halos then the CB representing plus one simply starts at unaligned access offset starting location plus two, whereas the CB representing minus one starts at the unaligned access starting location.

Initially it looked like this would be easiest to implement by the data mover core, modifying the field in the appropriate CB structure. However, data mover and compute cores maintain separate copies of this structure, meaning that changes made by the data mover to CB pointers are not visible by the compute core. Furthermore, we obtained a linking error when including the \emph{cb\_interface} array in the compute core code as the definition of this can not be found.

Consequently, we added an additional API call into tt-metal's \emph{cb\_api.h} header file, \emph{cb\_set\_rd\_ptr}, which instructs the unpack compute core to call into \emph{llk\_set\_read\_ptr}, and this is a function that we added to the Grayskull specific part of the SDK to undertake the actual pointer assignment to the read field. We pass the local memory buffer as a compile argument to the compute kernel and the compute cores modify their read pointers once the \emph{cb\_wait\_front} call completes for a CB.

\section{Performance and energy efficiency comparison}
\label{sec:performance}

We scaled up the kernel across the e150's 108 worker Tensix cores, adopting a systolic array approach by decomposing in two dimensions across the cores. Each core is allocated a set of batches, and in this section we use a global problem size of 1024 by 9216 (9.4 million) BF16 elements.

\begin{table}[htb]
    \centering
    \caption{Performance and energy usage comparison for a problem size of 1024 by 9216 (9.4 million) BF16 elements, over 5000 iterations.}
    \label{tab:final_performance}
    \begin{tabular}{|c|ccc|cc|}
    \hline  
    \textbf{Type} & \textbf{Total} & \textbf{Cores in} & \textbf{Cores in} & \textbf{Performance} & \textbf{Energy}\\
     & \textbf{cores} & \textbf{Y} & \textbf{X} & \textbf{(GPt/s)} & \textbf{(Joules)}\\
      \hline
        CPU & 1 & - & - & 1.41 & 1657 \\
        CPU & 24 & - & - & 21.61 & 588 \\
        \hline
        e150 & 1 & 1 & 1 & 1.06 & 2094 \\
        e150 & 2 & 1 & 2 & 2.48 & 893 \\
        e150 & 4 & 1 & 4 & 2.92 & 744 \\
        e150 & 8 & 4 & 4 & 7.99 & 276 \\
        e150 & 32 & 8 & 4 & 9.20 & 240 \\
        e150 & 64 & 8 & 8 & 12.96 & 170 \\
        e150 & 72 & 8 & 9 & 17.26 & 128 \\        
        e150 & 108 & 12 & 9 & 22.06 & 110 \\   
        \hline
        e150 x 2 & 216 & 24 & 9 & 44.12 & 102 \\
        e150 x 4 & 432 & 48 & 9 & 86.75 & 108 \\
        
    \hline
    \end{tabular}
\end{table}

Table \ref{tab:final_performance} reports our results, where it can be seen that for this optimised code on one Tensix core we obtain performance of 1.06 GPt/s. This is 163 times greater than the performance of our initial version in Section \ref{sec:initial_jacobi}, much closer to the performance of a single CPU core, and fairly close to the maximum achievable performance of 1.387 Gpt/s when just enabling the compute component. We then scaled the number of Tensix cores, and over 108 workers on the entire e150 we slightly outperform the 24-core CPU but use around five times less energy. The reason for this energy usage pattern is that the power draw of the e150 is roughly constant, between 50 and 55 Watts, regardless of the number of Tensix cores in use. 


The machine that we are using is equipped with four e150 cards and we therefore undertook an experiment scaling up across these cards. Strictly speaking this will not provide the correct answer, as the cards cannot access each other's memory for the boundary conditions, although it could be routed through PCIe via the host this is not supported currently by tt-metal. However, the next generation Wormhole cards are directly interconnected and can access remote memory directly. Therefore it is interesting to explore the potential performance that this can deliver, and it can be seen that performance scales well across the cards, delivering around four times the performance of the Xeon Platinum CPU on 432 Tensix cores, although this does increase the overall power draw so that the energy usage is roughly similar and again around five times less than the CPU.

\section{Conclusions, recommendations and future work}
\label{sec:conclusions}

In this paper we have explored porting and optimising the Jacobi iterative method for solving Laplace's equation for diffusion in two dimensions to the Tenstorrent Grayskull. Whilst this is a simple solver, it is representative of a much wider class of stencil based algorithms that are ubiquitous in HPC. We have demonstrated the key considerations in obtaining optimal performance for this code on the Grayskull, ultimately obtaining comparable performance to the CPU but at five times less energy usage on one e150. However, it should be highlighted that given the capabilities of the hardware, the CPU is running in FP32 whereas the Grayskull uses BF16.

It was our initial hypothesis that, by separating data movement from compute, the Grayskull could provide improved flexibility and performance for HPC codes. Based upon the work detailed in this paper we believe that this is the case, however have demonstrated that to obtain best performance one must carefully construct their DRAM memory accesses and avoid memory copies. Consequently, more flexibility around these components in the API would be welcome, for instance enabling CBs to alias local memory and the ability to provide a map to the read and write NoC calls that instruct the routers/DMA engines how to pack or unpack data. These would provide increased flexibility around memory accesses whilst avoiding the bottlenecks that we have identified.

We are now looking at more complex stencil algorithms, such as atmospheric advection, on the Grayskull and intend to explore porting our approach to the Wormhole card which, with support for FP32 by the FPU will enable increased precision, along with the ability to connect the cards to explore scaling up in more detail. We might also be able to obtain improved scaling across the Tensix cores by first copying the domain into local SRAM and operating from there, although this would limit the size of the domain and require direct neighbour to neighbour communications.

\section*{Acknowledgement}
This work has been funded by the ExCALIBUR H\&ES RISC-V testbed. The CPU runs in this paper ran on NextGenIO which which funding from the EU Horizon 2020 research and innovation programme under grant agreement No 671591. We would like to thank Tenstorrent, and especially Eric Duffy, for kindly providing access to the Grayskull cards used throughout this work and for their suggestions and advice during development. For the purpose of open access, the author has applied a Creative Commons Attribution (CC BY) licence to any Author Accepted Manuscript version arising from this submission.

\bibliographystyle{IEEEtran}
\bibliography{references.bib}
\end{document}